\documentclass[twocolumn,jcp,showpacs,preprintnumbers,amsmath,amssymb,floatfix,superscriptaddress]{revtex4-1}
\usepackage{graphicx}
\usepackage{amsmath}
\usepackage{amssymb}
\usepackage{color}
\usepackage{cases}

\begin{document}

\title{Tunneling conductivity in composites of attractive colloids}

\author{B. Nigro}\email{biagio.nigro@epfl.ch}\affiliation{LPM, Ecole Polytechnique F\'ed\'erale de
Lausanne, Station 17, CP-1015 Lausanne, Switzerland}
\author{C. Grimaldi}\email{claudio.grimaldi@epfl.ch}\affiliation{LPM, Ecole Polytechnique F\'ed\'erale
de Lausanne, Station 17, CP-1015 Lausanne, Switzerland}
\author{M. A. Miller}\affiliation{University Chemical Laboratory, Lensfield Road,
Cambridge CB2 1EW, United Kingdom}
\author{P. Ryser}\affiliation{LPM, Ecole Polytechnique F\'ed\'erale de
Lausanne, Station 17, CP-1015 Lausanne, Switzerland}
\author{T. Schilling}\affiliation{Universit\'e du Luxembourg, 162 A, avenue de la
Fa\"iencerie, L-1511 Luxembourg}



\begin{abstract}
In conductor-insulator nanocomposites in which conducting fillers are dispersed
in an insulating matrix the electrical connectedness is established by interparticle tunneling
or hopping processes. These systems are intrinsically non-percolative and a coherent description of
the functional dependence of the conductivity $\sigma$ on the filler properties, and in particular
of the conductor-insulator transition, requires going beyond the usual continuum percolation approach
by relaxing the constraint of a fixed connectivity distance.
In this article we consider dispersions of conducting spherical particles which are connected to all
others by tunneling conductances and which are subjected to an effective attractive square well potential.
We show that the conductor-insulator transition at low contents $\phi$ of the conducting fillers does not
determine the behavior of $\sigma$ at larger concentrations, in striking contrast to what is predicted
by percolation theory. In particular, we find that at low $\phi$ the conductivity is governed almost
entirely by the stickiness of the attraction, while at larger $\phi$ values $\sigma$ depends mainly
on the depth of the potential well. As a consequence, by varying the range and depth of the potential
while keeping the stickiness fixed, composites with similar conductor-insulator transitions may display
conductivity variations of several orders of magnitude at intermediate and large $\phi$ values.
By using a recently developed effective medium theory and the critical path approximation we explain this
behavior in terms of dominant tunneling processes which involve interparticle distances spanning different
regions of the square-well fluid structure as $\phi$ is varied. Our predictions could be tested in experiments
by changing the potential profile with different depletants in polymer nanocomposites.
\end{abstract}


\maketitle

\section{Introduction}
\label{sec:intro}
The challenge of understanding the electrical transport properties in conductor-insulator
composites is central for conceiving and designing new composite materials with unique electrical
properties, and has fueled ongoing research ranging from fundamental statistical physics
to the more applied materials science and nanotechnology.
In this respect, recent years have seen remarkable progress in the design and synthesis of
polymer nanocomposites  with controllable structural parameters,
so as to combine advantageous properties of the insulating polymer matrix
(flexibility, light weight, transparency, \textit{etc.}) with appropriate levels of electrical
conductivity $\sigma$.
For example, high aspect-ratio conducting fillers, such as carbon fibers and nanotubes, graphite
platelets and graphene sheets, as well as segregated dispersions of carbon-black particles, have been shown
to reduce strongly the loadings $\phi$ needed to achieve conducting composites \cite{Bauhofer2009,Kuilla2010,Grunlan2001}.
The interplay of multicomponent fillers has been recently demonstrated to have a strong influence on transport
properties \cite{Kyrylyuk2011}.  Other studies have evidenced the role of depletants \cite{Vigolo2005} and of
shear forces \cite{Kharchenko2004} in the formation and structure of the conducting network of carbon nanotubes,
thus underlining the importance of the coupling between the conducting and insulating phases \cite{Hermant2009},
and how this can be used to tune the overall composite conductivity.

On the theoretical side, the general understanding of the effects of the filler properties on $\sigma$
has been traditionally based on the continuum percolation theory of transport.
The premise behind this approach is that there exists a fixed
inter-particle distance beyond which two particles are electrically disconnected, thereby leading to a
well-defined percolation threshold $\phi_c$ for the connected particles which marks the transition
between the conducting (at $\phi>\phi_c$) and the insulating (at $\phi <\phi_c$) regimes of the
composite \cite{Stauffer1994,Sahimi2003}.
In the vicinity of $\phi_c$, the resulting percolation conductivity is thus predicted to follow the
power-law relation
\begin{equation}
\label{sigmaperc}
\sigma_{\rm perc}\simeq \sigma^0_{\rm perc}(\phi-\phi_c)^t,
\end{equation}
where $t\simeq 2$ is the universal transport exponent for three-dimensional percolating systems.
Equation \eqref{sigmaperc} is customarily used to interpret the $\phi$-induced conductor-insulator
transition observed in real composite materials, which are thus viewed
as truly percolating systems, although the values of $t$ extracted from experiments often deviate
from universality \cite{Bauhofer2009,Vionnet2005}. Due to the percolation hypothesis, most theoretical
efforts have been concentrated on the calculation of the percolation threshold $\phi_c$ and of
its dependencies on the filler particle shapes \cite{Balberg1984,Otten,Chatterjee2010},
dispersions \cite{Kusy1977,Johner2009}, and filler-filler and filler-matrix
interactions \cite{Chiew1983,Bug1985,Schilling2007,Kyrylyuk2008}.

In polymer nanocomposites, or more generally in colloidal-like dispersions of conducting
fillers in insulating matrices, the electron transfer between particles is mediated by tunneling
or hopping processes, which implies that the conductance between any two particles decays
continuously with the inter-particle distance [see Eq.~\eqref{tunnel} below].
In this case, no fixed connectivity length can be unambiguously identified and, consequently,
the premise for the existence of a percolation transition, and for the
validity of Eq.~\eqref{sigmaperc}, is unjustified for this class of composites.
This poses the problem of understanding the observed filler dependencies of $\sigma$ without
relying on the percolation hypothesis.

\begin{figure*}[t!]
\begin{center}
\includegraphics[scale=0.63,clip=true]{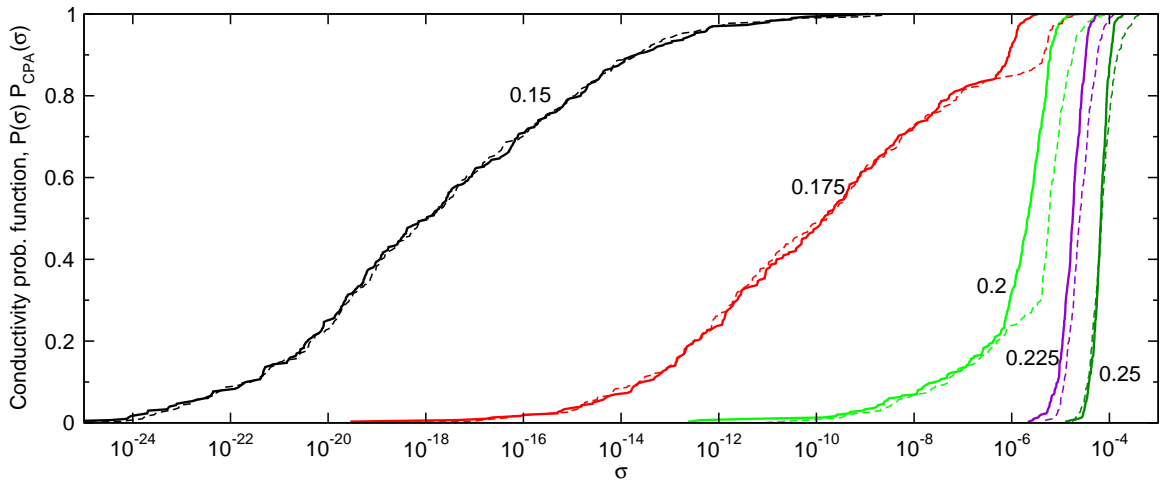}
\caption{(Color online) Conductivity probability function $P(\sigma)$ (thick solid lines) for
various values of the volume fraction $\phi$ and for a square-well potential with $\lambda=1.05$ and
$\tau=0.2$. Here the tunneling factor is $\xi/D=0.01$ and the number of conducting spheres is fixed
at $N=2000$. The thin dashed lines are the $P_{\rm CPA}(\sigma)$ curves obtained from $P(\delta)$
of Fig.~\ref{fig6} and by using $\sigma=\sigma_0\exp\!(-2\delta/\xi)$ with $\sigma_0=0.09$.}\label{fig1}
\end{center}
\end{figure*}

Recently, the problem of finding the functional dependence of $\sigma$ for colloidal composites
without imposing any \textit{a priori} fixed connectivity distance has been tackled by considering
explicitly the electron tunneling conduction between any two particles in the
composite \cite{Ambrosetti2010a,Ambrosetti2010b,Nigro2011}.
One important outcome was the realization that, despite the absence of a real percolation threshold,
the transition from the conducting to the insulating regimes could still be characterized by a characteristic
concentration $\phi_p$ below which $\sigma$ is dominated by the (small but finite) conductivity $\sigma_p$
of the insulating matrix. In particular, by considering equilibrium dispersions of hard spheroidal particles,
$\phi_p$ was shown to decrease with the particle aspect-ratio \cite{Ambrosetti2010a}, in accord with
the general trend observed in composites with fibrous and plate-like fillers.

In this article we study by theory and simulations the functional dependence of $\sigma$ when, in addition
to steric interactions, the conducting colloidal particles are subjected to mutual
attraction modeled by a square well potential. We show that attraction, when combined with tunneling, gives rise to unexpected
features which could not be anticipated by the usual continuum percolation theory.
Specifically, we find that the range and the depth of the attraction affect the conductivity
in very different ways according to whether $\phi$ is small or large. In particular, the
low-$\phi$ behavior of $\sigma$, and so the crossover $\phi_p$ to the insulating regime, turns out to
be almost entirely governed by the overall stickiness of the attraction,
while the conductivity at $\phi>\phi_p$ is found to change by several orders of magnitude for
different potential profiles even if the stickiness is kept constant.
This kind of behavior is completely missed if the conductor-insulator crossover
is treated as a true percolation transition occurring at the critical concentration $\phi_c$, because
in this case the $\phi$ dependence of the percolation conductivity $\sigma_{\rm perc}$ of Eq.~\eqref{sigmaperc}
is basically determined by $\phi_c$ alone.

As shown in the following, in which a recently developed effective medium theory and the critical
path approximation are used to interpret the simulation results, the effects of the attractive
potential on $\sigma$ can be rationalized in terms of a $\phi$-dependent dominant distance between
the fillers which governs the composite conductivity. This correspondence establishes a direct connection
between the functional dependence of $\sigma$ and the structure of the square well fluid, which
could be in principle exploited in the synthesis of real nanocomposite materials with optimized
conduction characteristics.

\section{Model}
\label{sec:model}

Attractive forces between filler particles in polymer nanocomposites may arise
from several mechanisms such as van der Waals forces or depletion
interactions induced by non-adsorbing polymers or surfactant micelles \cite{Lekkerkerker2011}.
In the latter case, the size and concentration of the depletants control respectively the range and depth of the
attractive interaction, which can be modeled by a suitable effective potential.
Here, we represent the conducting fillers by hard-core spheres of diameter $D$
and for any two spheres centered at $\mathbf{r}_i$ and $\mathbf{r}_j$, we model the
attraction by a square-well potential $u(r_{ij})$ of the form:
\begin{equation}
\label{umodel}
u(r_{ij})=\left\{
\begin{array}{ll}
+\infty & \mbox{ for $r_{ij}\leq D$} \\
-\epsilon & \mbox{ for $D< r_{ij}\leq \lambda D$} \\
0 & \mbox{ for $r_{ij}>\lambda D$}
\end{array}\right.
\end{equation}
where $r_{ij}=|\mathbf{r}_i -\mathbf{r}_j|$ and $\lambda D$ ($\lambda\geq 1$)
is the range of attraction. In the following, we shall vary both $\lambda$ and $\epsilon$
so as to consider the hard sphere (HS) case ($\epsilon=0$), the short-range attraction
regime $\epsilon\neq 0$ and $\lambda\leq 1.25$, and the adhesive hard sphere (AHS)
limit \cite{Baxter1968}, which is obtained by taking $\lambda\rightarrow 1$ and
$\epsilon\rightarrow\infty$ with
\begin{equation}
\label{tau}
\tau=\frac{\lambda\exp(-\epsilon^*)}{12(\lambda-1)}
\end{equation}
constant, where $\epsilon^*=\epsilon/kT$ and $kT$ is the thermal energy.
$\tau$ is an inverse measure of particle stickiness which governs, together with the concentration,
the phase behavior of short-range square-well fluids \cite{Noro2000,Foffi2006}.
In the following, we limit our study to values
of $\tau$ above the critical point for gas-liquid-like phase separation \cite{Largo2008}.

We describe the electron transport processes by considering each sphere as being electrically
connected to all others through inter-particle tunneling conductances of the form
\begin{equation}
\label{tunnel}
g(r_{ij})=g_0\exp\!\left[-\frac{2(r_{ij}-D)}{\xi}\right],
\end{equation}
where $\xi$ is the tunneling decay length, which is independent of the range of the attraction $\lambda$.
In writing Eq.~\eqref{tunnel} we assume that the size $D$ of the conducting particle and the
temperature are large enough to neglect charging energy effects and Coulomb interactions.
Furthermore, we have ignored any dependence on $r_{ij}$ of the prefactor $g_0$, which we set
equal to $1$.

\section{Numerical results for the conductivity}
\label{sec:numerical}
We performed standard Metropolis Monte Carlo simulations to find equilibrium dispersions of $N=2000$
particles in a cubic box of side length $L$, which was changed to obtain different values
of the volume fraction $\phi=\pi N D^3/6L^3$. A dedicated Monte Carlo algorithm \cite{Miller2004}
was used for the AHS limit. For all the $\phi$ values
considered and for each potential profile we obtained $N_R=300$ independent equilibrium
configurations of the system.

To obtain the composite conductivity $\sigma$, we constructed for each of the $N_R$ configurations
a tunneling resistor network with inter-particle conductances $g(r_{ij})$ given by Eq.~\eqref{tunnel}.
In order to reduce computational times, we exploited the exponential decay of Eq.~\eqref{tunnel} by neglecting
contributions from particles sufficiently far apart so as to reduce the number of connected particles
in the network. To this end, we introduced an adjustable
cut-off length $\delta_{\rm max}$ such that the bond conductances for $\delta_{ij}>\delta_{\rm max}$
can be safely removed from the network without altering $\sigma$ \cite{Ambrosetti2010a,Nigro2011}.
Finally, we solved this reduced network using a combination of numerical decimation and preconditioned
conjugate gradient algorithms as described in Ref.~\onlinecite{Nigro2011}. From the resulting
network conductance $G$ the dimensionless conductivity $\sigma$ follows from $\sigma=GD/L$.

In Fig.~\ref{fig1} we show some representative examples of the cumulative distribution function (CDF) $P(\sigma)$
of the conductivity (thick solid lines) obtained from all $N_R$ realizations of the network for
$\lambda=1.05$, $\tau=0.2$, and for different values of the volume fraction $\phi$. In all cases the
tunneling decay length has been set equal to $\xi/D=0.01$.
Overall, the rise from $0$ to $1$ of $P(\sigma)$ becomes more gradual as $\phi$ decreases,
which comes from keeping the number of particles $N$ fixed. Additionally, for $\phi=0.175$
and $0.2$ the CDF displays a sudden increase at $\sigma\simeq 0.5\times 10^{-7}$.
As discussed in more detail in Sec.~\ref{sec:CPA}, this latter feature can be traced back to the discontinuity of the
radial distribution function of the square-well fluid at interparticle distances $r=\lambda D$.
Since the numerical procedure for solving the tunneling network is rather time consuming,
we have not attempted to study finite size-effects on $P(\sigma)$ systematically. Instead, we have chosen
to extract the overall network conductivity from the condition $P(\sigma)=1/2$ applied to the CDF
computed for $N=2000$. As shown in Sec.~\ref{sec:CPA}, where we study the conductivity within the critical path
approximation for which a finite-size analysis is feasible, this criterion provides a robust estimate of
the conductivity for infinite systems.

\begin{figure*}[t!]
\begin{center}
\includegraphics[scale=0.4,clip=true]{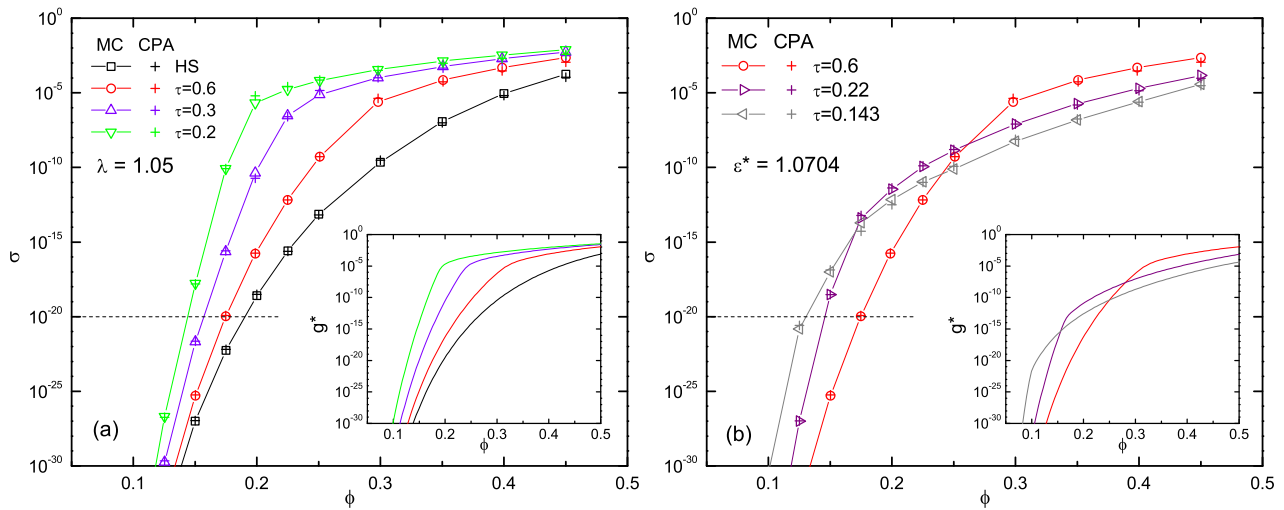}
\caption{(Color online) Network conductivity $\sigma$ as a function of the volume fraction $\phi$
obtained from MC calculations (open symbols) and from CPA (plus signs). (a) $\sigma$ for well range
fixed at $\lambda=1.05$ and different well depths $\epsilon^*$ parametrized
by $\tau$ of Eq.~\eqref{tau}. (b) $\sigma$ for well depth fixed at $\epsilon^*=1.0704$ and
$\lambda=1.05$, $1.15$, and $1.25$.
Insets: corresponding EMA conductance $g^*$.}\label{fig2}
\end{center}
\end{figure*}

In the main panels of Fig.~\ref{fig2} we show the network conductivity $\sigma$ (open symbols)
as a function of $\phi$ for $\xi/D=0.01$ and for different parameters $\lambda$ and $\epsilon^*$ of the square-well
potential. For comparison, the HS limit is plotted in Fig.~\ref{fig2}(a) as open squares.
By inspection of the two panels of the figure we see that the $\phi$-dependence of $\sigma$
strongly depends on the way in which the potential profile is changed. Namely, in Fig.~\ref{fig2}(a)
the conductivity is steadily enhanced for all $\phi$ values when the well depth is increased while keeping
the range parameter fixed at $\lambda=1.05$, leading for example at $\phi=0.25$ to an enhancement of up to
$9$ orders of magnitude compared to the HS limit.
Conversely, as shown in Fig.~\ref{fig2}(b), when the potential depth is kept as fixed ($\epsilon^*=1.0704$) but its range varies
from $\lambda=1.05$ to $\lambda=1.25$ the conductivity at low $\phi$ is enhanced, while at intermediate and large
$\phi$ it is diminished by about three orders of magnitude.

The different functional dependencies of $\sigma$ in Figs.~\ref{fig2}(a) and (b) have important
consequences in conjunction with the conductor-insulator transition of composites.
Indeed, although we have considered conducting particles dispersed in a perfectly insulating medium,
in real polymer composites the conductivity $\sigma_p$ of the polymeric matrix is small but finite
at nonzero temperatures, being typically $10^{-13}-10^{-18}$ S/cm at room temperature.
Hence, despite the fact that the tunneling conductivity drops all the way to zero as $\phi\rightarrow 0$,
the conductivity of the complete system (particles and polymer) is limited from below by $\sigma_p$ \cite{NoteSec3}.
One can thus identify a crossover point separating tunneling-dominated from polymer-dominated conductivity
with the volume fraction $\phi_p$ below which $\sigma$ matches that of the polymer matrix, as is
generally done in the analysis of experimental conductivity data of real nanocomposites.
As schematically shown in the main panels of Fig.~\ref{fig2}, where the horizontal dashed line represents $\sigma_p$
fixed for illustrative purposes at $10^{-20}$,
we can estimate $\phi_p$ as the intersection point between the calculated tunneling conductivity
and $\sigma_p$ \cite{Ambrosetti2010a}. In this way we find that as the stickiness of the square-well is enhanced
(\emph{i.e.}, $\tau$ is diminished) $\phi_p$ is systematically lowered. However, it is evident by comparing
Fig.~\ref{fig2}(a) with Fig.~\ref{fig2}(b), that the behavior of $\sigma$ in the conducting regime $\phi>\phi_p$
is not determined by $\phi_p$ alone.  For example the curves for $\tau=0.2$ in Fig.~\ref{fig2}(a) and for
$\tau=0.22$ in Fig.~\ref{fig2}(b) have similar $\phi_p$ but very different conductivities at larger $\phi$.

\begin{figure}[t!]
\begin{center}
\includegraphics[scale=0.4,clip=true]{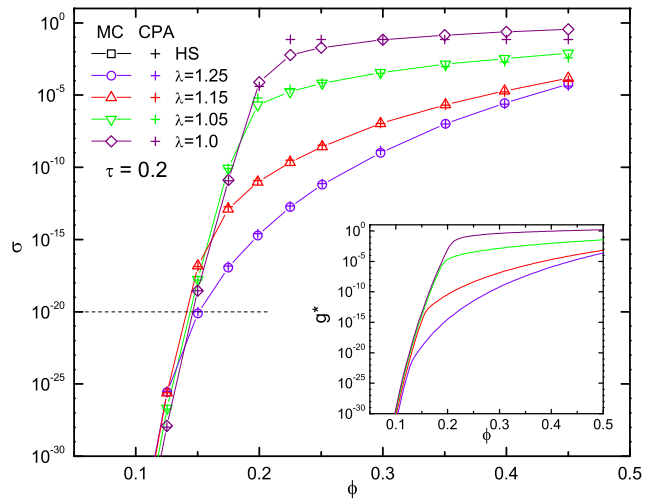}
\caption{(Color online) Network conductivity $\sigma$ as a function of $\phi$ obtained from MC
calculations (open symbols) and from CPA (plus signs) for different well widths and depths with
fixed $\tau=0.2$. Inset: corresponding EMA conductance $g^*$.}\label{fig3}
\end{center}
\end{figure}

This latter feature is illustrated even more strikingly in Fig.~\ref{fig3} where $\sigma$ is plotted for different values of
$\epsilon^*$ and $\lambda$ such that the stickiness parameter $\tau$ remains constant.
Even though $\phi_p$ is basically unchanged in going from $\lambda=1.25$ to
the AHS limit at $\lambda=1$, the conductivity increases by $5$ to $10$ orders of magnitude above
the HS case at fixed $\phi$ in the whole $\phi>\phi_p$ range. Note that a similar behavior is
also found if we change the value of $\sigma_p$ provided that it remains sufficiently small.
For example if in Fig.~\ref{fig3} we consider $\sigma_p=10^{-15}$, the curves
for $\lambda\leq 1.15$ would still have equal $\phi_p$, though slightly larger than the previous case,
but very different $\sigma$ at larger $\phi$.

Our general observation that in square-well fluids of conducting spheres the conductivity
for $\phi>\phi_p$ is not determined by the position of the conductor-insulator crossover is in
striking contrast with the usual continuum percolation description. Indeed if the conductor-insulator
transition is viewed as a true percolation transition, then the corresponding conductivity would
follow Eq.~\eqref{sigmaperc} for $\phi\gtrsim \phi_c$, where $\phi_c$ is the percolation threshold,
and so the conductivity level in the conducting regime would be completely determined by $\phi_c$.
For the short-range potentials considered here $\phi_c$ is known to be reduced by enhanced
stickiness \cite{Bug1985}, in line with the behavior of $\phi_p$, and consequently, from
Eq.~\eqref{sigmaperc}, the percolation conductivity for $\phi\gtrsim \phi_c$ would also be
systematically enhanced. Conversely, different potentials with the same parameter $\tau$
would give similar percolation conductivity curves, in strong disagreement with our results of
Fig.~\ref{fig3}.

\begin{figure}[t!]
\begin{center}
\includegraphics[scale=0.4,clip=true]{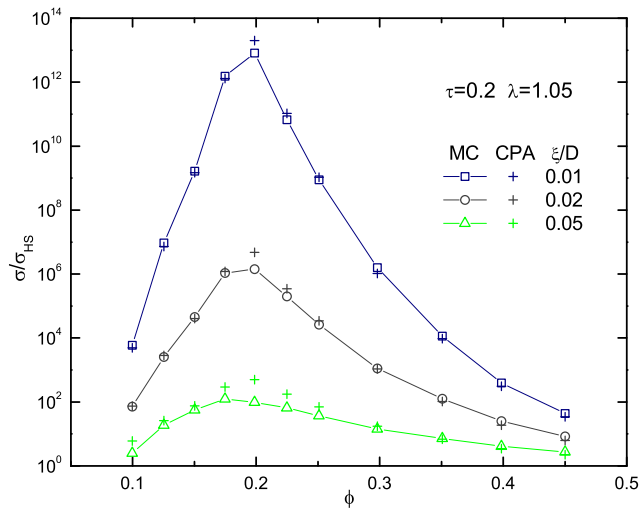}
\caption{(Color online) Enhancement factor of the conductivity for $\tau=0.2$ compared
to the conductivity in the HS limit for different values of the tunneling length $\xi/D$.}\label{fig4}
\end{center}
\end{figure}

The results shown in Figs.~\ref{fig2} and \ref{fig3} also have important practical consequences
in relation to the problem of controlling the conductivity of polymer nanocomposites by tuning
the attraction through, for example, the depletion interaction \cite{Vigolo2005}.
Indeed, from Figs.~\ref{fig2} and \ref{fig3}, it turns out that the best strategy
to obtain high levels of conductivity for a broader range of filler concentrations is by
choosing potential profiles that are very deep and short-ranged (but of course not strong enough
to induce phase separation). For depletion induced attraction, this corresponds to having a large
concentration of depletants with small sizes compared to those of the conducting particles.

\begin{figure*}[t!]
\begin{center}
\includegraphics[scale=0.64,clip=true]{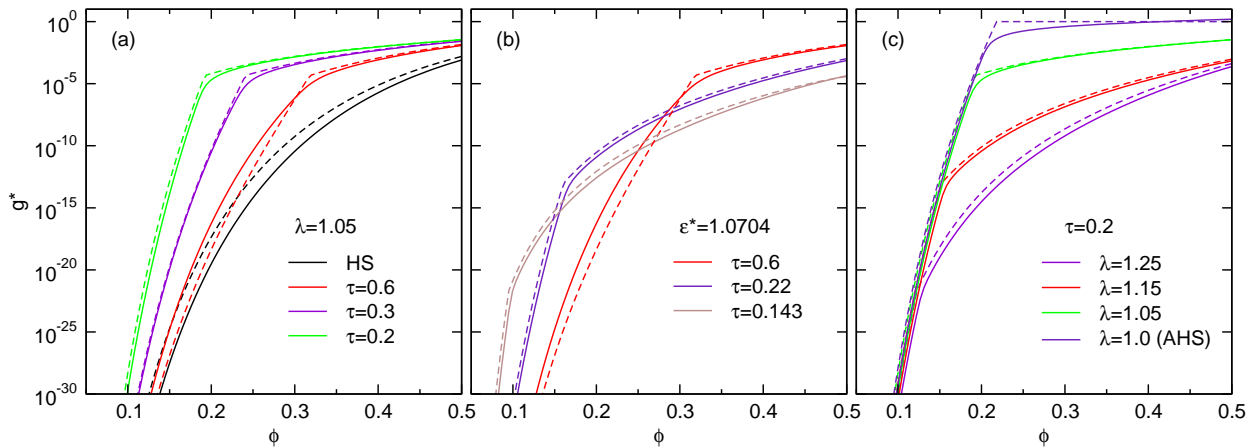}
\caption{(Color online) Comparison between the full numerical solution of the EMA equation \eqref{EMA}
(solid lines) and $g^*$ obtained from Eqs.~\eqref{EMA2} and \eqref{EMA3} (dashed lines). The parameters used
in (a), (b), and (c) are the same of those in Figs.~\ref{fig2}(a), \ref{fig2}(b), and \ref{fig3}, respectively.}\label{fig5}
\end{center}
\end{figure*}

We conclude this section by addressing the role of the tunneling decay length $\xi$ in
the overall functional dependence of $\sigma$. In Fig.~\ref{fig4} we plot $\sigma$
for $\tau=0.2$ and $\lambda=1.05$ in units of the HS conductivity for $\xi/D=0.01$, $0.02$, and $0.05$.
As expected, the strong enhancement obtained for $\xi/D=0.01$ is systematically reduced as $\xi/D$ increases.
For example at $\xi/D=0.05$ only a maximum $10^2$ fold enhancement with respect to the HS limit is achieved.
This is because for large $\xi/D$, two particles having separation less or greater than
the potential range have similar tunneling probabilities, and so the short-range structure of the square-well
fluid is smeared out. If we consider that, depending on the material properties, the tunneling decay length
ranges from a fraction of a nanometer to a few nanometers, from the results of Fig.~\ref{fig3}
we expect that particle sizes of tens to hundreds of nanometers are optimal for $\sigma$ to be strongly
enhanced by increased attraction.

\section{Effective medium theory}
\label{sec:EMA}
The numerical results of Sec.~\ref{sec:numerical} and the effects of attraction on the
functional dependence of $\sigma$ find a more complete understanding by
employing the effective medium approximation (EMA) introduced in
Refs.~\onlinecite{Ambrosetti2010b,Grimaldi2011}. Briefly, this EMA amounts to
replacing each conductance $g(r_{ij})$ of Eq.~\eqref{tunnel}
by an effective value, $\bar{g}$, which is independent of the inter-particle distance $r_{ij}$.
By requiring that the resulting effective and fully-connected network has the same average resistance
as the original, and by considering only two-site clusters \cite{Grimaldi2011}, the following self-consistent
equation is found \cite{Ambrosetti2010b,Grimaldi2011}:
\begin{equation}
\label{EMA}
24\phi\int_0^\infty\! dx\, x^2 g_2(x)\left[\frac{1}{g^*\exp[2D(x-1)/\xi]+1}\right]=2,
\end{equation}
where $x=r/D$, $g_2(x)$ is the radial distribution function (RDF) for the conducting spheres, and
$g^*=N\bar{g}/2$ is the two-point conductance of the effective $N$-node network.

We have numerically solved
Eq.~\eqref{EMA} for $g^*$ using the quasi-analytical model RDF proposed in Ref.~\onlinecite{Yuste1994} for the
square-well fluid, which reduces to the Percus-Yevick approximation for both the HS and AHS limits and
is in rather good agreement with the MC results of the RDF for $\lambda< 1.3$ \cite{Yuste1994,Largo2005}.
Although $g^*$ is a two-point conductance (which is the meaningful quantity for a complete network of
identical resistors) rather than a conductivity,
the EMA results shown in the insets of Figs.~\ref{fig2} and \ref{fig3} and in Fig.~\ref{fig5}
(solid lines) closely follow the conductivity
behavior obtained by the full numerical solution of the tunneling network. Using Eq.~\eqref{EMA},
we can thus explain the effect of attraction on $\sigma$ by reasoning simply in terms of
$g_2(x)$ and of its dependence on the square-well potential.
To this end, let us rewrite the EMA conductance as
\begin{equation}
\label{EMA2}
g^*=\exp\!\left[-2D(x^*-1)/\xi\right],
\end{equation}
where we have introduced a characteristic EMA distance $x^*$. By noting that for $\xi/D\ll 1$
the term in square brackets in Eq.~\eqref{EMA} reduces to the step function $\theta(x^*-x)$,
$x^*$ is found as the solution of the following equation:
\begin{equation}
\label{EMAzeta}
24\phi\int_0^{x^*}\! dx\, x^2 g_2(x)=2,
\end{equation}
where the left hand side gives the number of particle centers having distances less than $x^*$
from a particle at the origin. To proceed further, we note that for sufficiently narrow and deep
potential wells the RDF can be approximated by
\begin{equation}
\label{RDF}
g_2(x)=\theta(x-1)[g_2(1^+)\theta(\lambda-x)+1],
\end{equation}
where $g_2(1^+)$ is the RDF at contact, and from Eq.~\eqref{EMAzeta} we obtain finally
\begin{subnumcases}{\label{EMA3}  x^*=}
\left[\frac{1}{4\phi g_2(1^+)}+1\right]^{\frac{1}{3}}  \mbox{ for $x^*<\lambda$,}\label{EMA3a}\\
\left[\frac{1}{4\phi}-g_2(1^+)(\lambda^3-1)+\lambda^3\right]^{\frac{1}{3}}  \mbox{ for $x^*>\lambda$.}
\,\,\,\,\,\,\,\,\,\,\label{EMA3b}
\end{subnumcases}
As shown in Fig.~\ref{fig5}, $g^*$ obtained from Eqs.~\eqref{EMA2} and \eqref{EMA3} (dashed lines) is
in overall good agreement with the solutions of Eq.~\eqref{EMA} with the full RDF of
Ref.~\onlinecite{Yuste1994}. Furthermore, Eq.~\eqref{EMA3}
explicitly introduces two different regimes according to whether the EMA distance $x^*$ is
less or greater than $\lambda$, which physically corresponds to whether the dominant tunneling
processes occur between particles with separation less or greater than the well range.
The two regimes $x^*>\lambda$ and $x^*<\lambda$ identify two different regions, respectively
$\phi<\phi^*$ and $\phi>\phi^*$, where $\phi^*$ is a characteristic concentration (not to be confused
with $\phi_p$) obtained from the condition $x^*=\lambda$ applied to Eq.~\eqref{EMA3}, \emph{i.e.},
$4\phi^* g_2(1^+)(\lambda^3-1)=1$.
Since $g_2(1^+)$ scales essentially as $\exp(\epsilon^*)$ \cite{Yuste1994}, $\phi^*$
can be approximated by $\tau /\lambda^2$ for $\lambda-1$ and $\tau$ small.

\begin{figure*}[t!]
\begin{center}
\includegraphics[scale=0.63,clip=true]{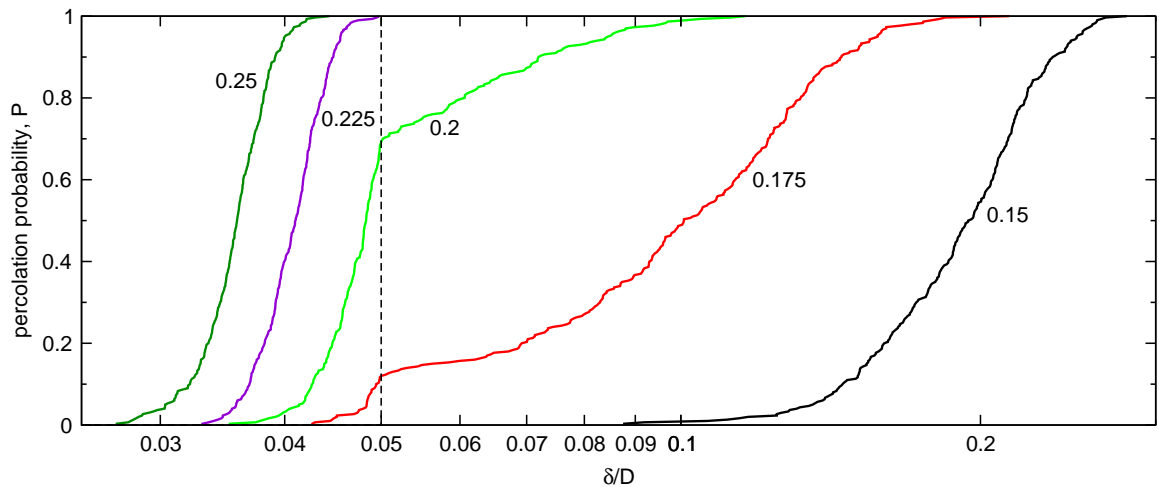}
\caption{(Color online) Percolation probability as a function of the connectivity distance $\delta$ for
various values of the volume fraction $\phi$ and for a square-well potential with $\lambda=1.05$ and
$\tau=0.2$ (i.e. $\epsilon^*=2.169$). The number of conducting spheres is fixed at $N=2000$.
The vertical dashed line indicates $\delta/D=\lambda-1$.}\label{fig6}
\end{center}
\end{figure*}

From Eq.~\eqref{EMA3a} and $g_2(1^+)\approx\exp(\epsilon^*)$ we thus see that the conduction at
large densities ($\phi>\phi^*$) is mainly governed by the well depth, which is indeed expected since in this
regime $x^*<\lambda$. In contrast, from Eq.~\eqref{EMA3b}, $g^*$ turns out to be
affected by both $\epsilon^*$ and $\lambda$ at low densities ($\phi<\phi^*$) because
the relevant tunneling distances span the entire well range (\emph{i.e.}, $x^*>\lambda$).
In this regime and for sufficiently narrow wells, $g_2(1^+)(\lambda^3-1)-\lambda^3$
reduces simply to $1/4\tau-1+\mathcal{O}([\lambda-1]/\tau)$, which explains the equal asymptotic
behaviors for $\phi\rightarrow 0$ of Fig.~\ref{fig3} as $\epsilon^*$ and $\lambda$ are changed but
$\tau$ is held fixed.

From the low-density approximation derived above, it is also possible to deduce the conductor-insulator
crossover point $\phi_p$ by introducing a two-point conductance $g_p\ll 1$ of the polymeric matrix,
which represents the EMA equivalent of $\sigma_p$ introduced in Sec.~\ref{sec:numerical}. Thus by
requiring that Eq.~\eqref{EMA2} coincides with $g_p$, and using Eq.~\eqref{EMA3b} in the narrow
well limit, we find for $(\xi/2D)\ln(1/g_p)\ll 1$
\begin{equation}
\label{phic}
\phi_p \approx \frac{\tau}{1+6(\xi\tau/D)\ln(1/g_p)}.
\end{equation}
Besides confirming that the structure of the square-well fluid affects $\phi_p$ only
through the stickiness, Eq.~\eqref{phic} also makes it explicit that the locus of the
conductor-insulator transition depends (although only logarithmically) on the insulating
matrix through $g_p$, in contrast to the usual continuum percolation theory which predicts
a percolation threshold $\phi_c$ independent of the matrix conductivity \cite{Ambrosetti2010a}.

\section{Critical path approximation}
\label{sec:CPA}
Although colloidal dispersions of tunneling connected particles are intrinsically non-percolative
systems, concepts and quantities of percolation theory are nevertheless at the basis of the
critical path approximation (CPA) \cite{CPApapers}  which, as already shown for HS fluids of
conducting particles \cite{Ambrosetti2010a,Nigro2011}, can reproduce to a high accuracy
the tunneling conductivity behavior of composites.

As we have seen in Sec.~\ref{sec:EMA}, for sufficiently small $\xi/D$
the EMA conductance $g^*$ is dominated by a characteristic length, $x^*$, such that
the cumulative coordination number satisfies Eq.~\eqref{EMAzeta}. In a similar
way, CPA amounts to approximating the tunneling network conductivity by
\begin{equation}
\label{CPA}
\sigma\simeq \sigma_0\exp\!\left[-\frac{2\delta_c(\phi)}{\xi}\right],
\end{equation}
where $\sigma_0$ is a $\phi$-independent conductivity prefactor and
$\delta_c(\phi)=r_c(\phi)-D$ is a critical distance given by the shortest among the $\delta_{ij}=r_{ij}-D$
lengths such that the subnetwork defined by all distances $\delta_{ij}\leq \delta_c(\phi)$ forms a
percolating cluster. From Eq.~\eqref{CPA} we see that CPA expresses $\sigma$ in terms of
a critical connectivity distance, $\delta_c(\phi)$, which is a genuine percolation quantity. However,
contrary to the  continuum percolation approach with fixed connectivity length, in CPA
the critical connectivity changes as $\phi$ and the potential profile are varied.
This adaptation of the connectivity range compensates for the artificial basis of the sharp cutoff at $\delta_c$.

To calculate $\delta_c(\phi)$ we follow the route described in Ref.~\onlinecite{Nigro2011}.
Namely, for fixed $\phi$ and potential profile,  we coat each sphere with a concentric penetrable
shell of thickness $\delta/2$, and consider two spheres as connected if their penetrable shells overlap.
A clustering algorithm described in Ref.~\onlinecite{Nigro2011} allows computation of the spanning
probability $P(\delta)$, which is plotted in Fig.~\ref{fig6} for $\lambda=1.05$, $\tau=0.2$, $N=2000$,
and for different values of $\phi$. Note that there is a rather sudden change in the slope of $P(\delta)$
when $\delta/D$ crosses $\lambda-1$ (vertical dashed line) which is due to the discontinuity of the
square well potential at $r/D=\lambda$.

\begin{figure*}[t!]
\begin{center}
\includegraphics[scale=0.6,clip=true]{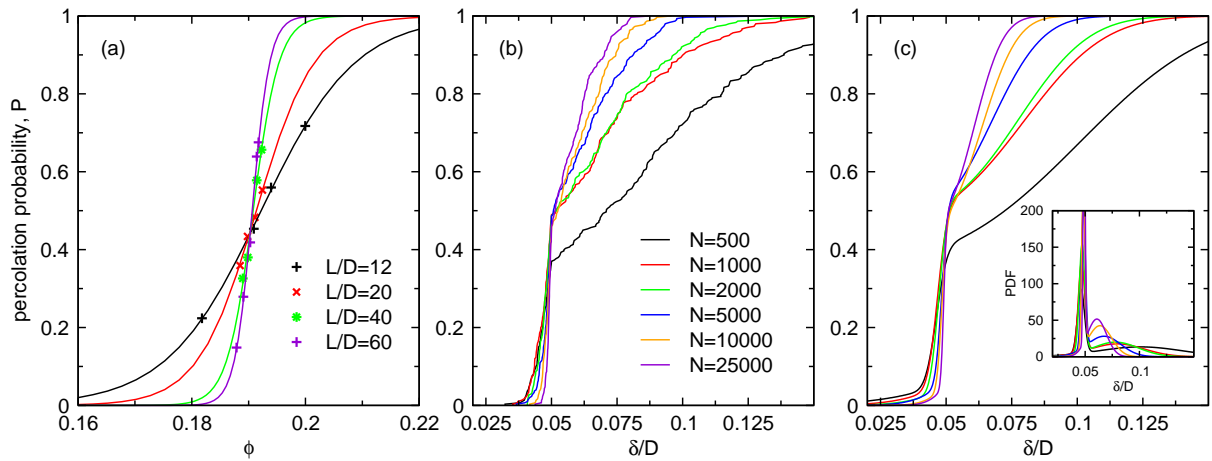}
\caption{(Color online) Percolation probability $P$ for $\lambda=1.05$ and $\tau=0.2$ (i.e. $\epsilon^*=2.169$).
(a): $P$ as a function of $\phi$ for $\delta/D$ fixed at $\lambda-1=0.05$ and for different sizes
$L$ of the cubic box. (b): $P$ as function of $\delta/D$ for $\phi=0.19$ and for different number
$N$ of spheres. (c): fits of the percolation probability of (b) obtained by using a linear
combination of two error functions. Inset: probability distribution function PDF obtained
from $dP(\delta)/d\delta$.}\label{fig7}
\end{center}
\end{figure*}

In order to define a suitable criterion to extract $\delta_c$
from the $P(\delta)$ curves for all parameter values used, and which is also insensitive to the feature
at  $\delta/D=\lambda-1$, we have carried out the finite-size scaling analysis of the type shown
in shown in Fig.~\ref{fig7}, where $P(\delta)$ is compared with the percolation probability $P(\phi)$
calculated as a function of $\phi$ for fixed $\delta$.
In Fig.~\ref{fig7}(a), we show the evolution of $P(\phi)$ as the system size $L/D$ increases for $\lambda=1.05$ and
$\delta/D=\lambda-1=0.05$. By using the finite-size scaling relation $\phi_c-\phi_c(L)\propto L^{-1/\nu}$,
where $\nu\simeq 0.88$ is the correlation length exponent and $\phi_c(L)$ is such that $P=1/2$,
we find that the percolation threshold as $L/D\rightarrow \infty$ is $\phi_c= 0.19035\pm 0.00001$.
In Fig.~\ref{fig7}(b) we plot $P(\delta)$ obtained for different numbers $N$ of particles by using $\phi=0.19$.
Note that as $N$ increases, the change of slope of $P(\delta)$ at $\delta/D=\lambda-1=0.05$
tends to decrease. This is more clearly seen in Fig.~\ref{fig7}(c), where we plot fits of $P(\delta)$ with
linear combinations of two error functions, and in the inset where we show the resulting probability
distribution function (PDF) $dP(\delta)/d\delta$, which is characterized by two peaks of different strengths.
If we use the criterion that $\delta_c(N)$ is given by $P=1/2$, we find from
$\delta_c-\delta_c(N)\propto N^{-1/3\nu}$ that $\delta_c/D=0.05066\pm 0.00007$ at $N\rightarrow\infty$,
practically coinciding with $\delta/D=0.05$ of Fig.~\ref{fig7}(a). Furthermore, for $N=2000$ the criterion
$P=1/2$ gives $\delta_c/D\simeq 0.0512$, which is only $1$\% larger than the critical distance at $N\rightarrow\infty$.

\begin{figure*}[t!]
\begin{center}
\includegraphics[scale=0.45,clip=true]{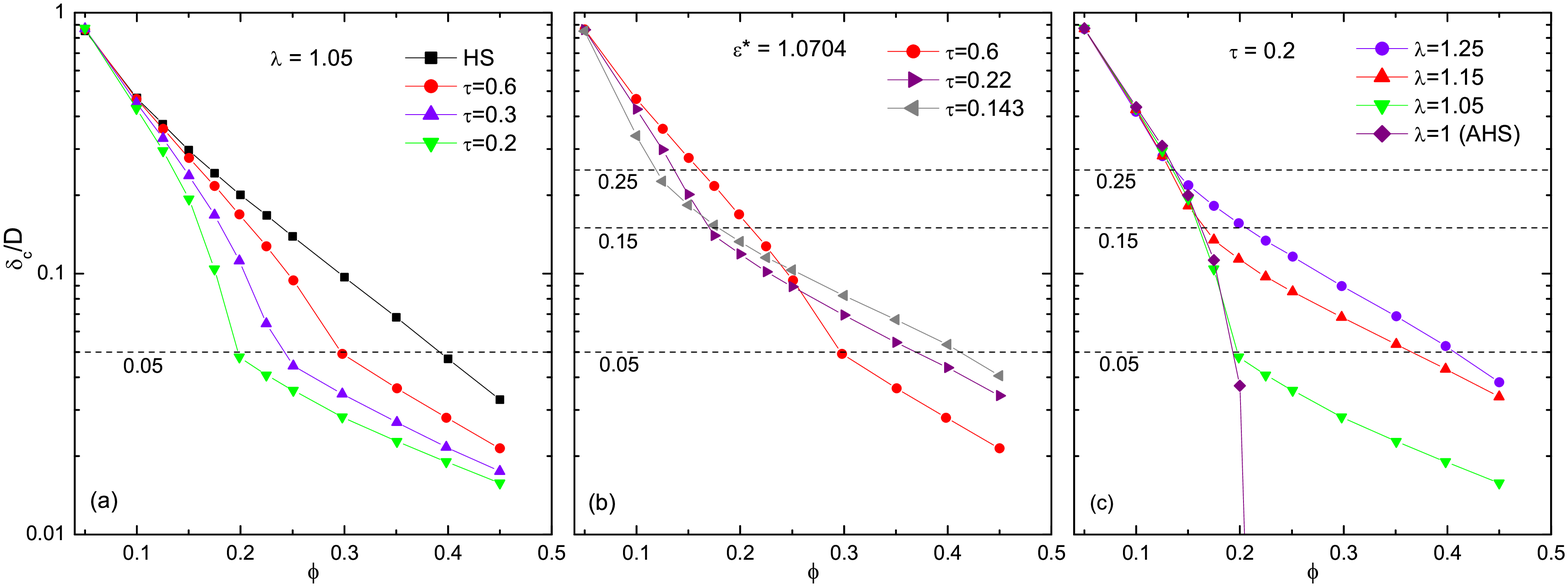}
\caption{(Color online) Calculated critical distance $\delta_c$ as a function of $\phi$ for
(a): $\lambda=1.05$ and different well depths; (b): $\epsilon^*=1.0704$ and $\lambda=1.05$,
$1.15$, and $1.25$; (c): $\lambda$ and $\epsilon^*$ varying with $\tau=0.2$ fixed.
The horizontal dashed lines identify $\delta_c/D=\lambda-1$. }\label{fig8}
\end{center}
\end{figure*}

From the finite-size study of Fig.~\ref{fig7}, and from other cases we have considered,
we thus see that $P(\delta_c)=1/2$ is a reliable criterion to find the critical distance even for
$N=2000$. When we compare in Fig.~\ref{fig1} the CDF of the network conductivity with $P_{\rm CPA}(\sigma)$ obtained from
$P(\delta$) by using the CPA expression $\sigma=\sigma_0\exp\!(-2\delta/\xi)$ with $\sigma_0=0.09$ (thin dashed lines),
we see that the shapes of the two probability functions are very similar, which justifies the criterion $P(\sigma)=1/2$
that we have adopted for the network conductivity results shown in Figs.~\eqref{fig2} and \eqref{fig3}.

Let us now discuss the conductivity dependence on $\phi$ and on the potential profile in
terms of the CPA formula \eqref{CPA}.
The calculated values of the critical distance are shown in Fig.~\ref{fig8} for the same
parameter values as Figs.~\ref{fig2} and \ref{fig3}. For all cases studied, $\delta_c$ is a monotonically
decreasing function of $\phi$, which is to be expected because larger particle concentrations bring
about shorter mean inter-particle distances.
However, the rate of decrease depends on whether $\delta_c/D$ is greater or less
than $\lambda-1$ (horizontal dashed lines in Fig.~\ref{fig8}), which reflects the existence of the two
different regions of the conductivity already discussed in Sec.~\ref{sec:EMA}.
In particular, as shown in Fig.~\ref{fig8}(c), when $\delta_c/D \gtrsim \lambda-1$ the critical distance
is governed solely by the particle stickiness and closely follows the AHS limit, in full correspondence
with the low-$\phi$ behavior of $\sigma$ of Fig.~\ref{fig3}.
Indeed when the $\delta_c(\phi)$ values of Fig.~\ref{fig8} are inserted in Eq.~\eqref{CPA},
we find that the resulting CPA conductivity (plus signs in Figs.~\ref{fig2} and \ref{fig3}) is in
excellent agreement with our MC values \cite{notesigma0}.

We therefore see that, in analogy with the results of Sec.~\ref{sec:EMA},  the conductivity changes
driven by attraction are due to the variations of the dominant tunneling distances, which are represented
by the effective connectivity length $\delta_c$. From Eq.~\eqref{CPA} we see also that in the presence
of a finite conductivity $\sigma_p$ of the insulating matrix, tunneling dominates the overall conductivity
as long as $\delta_c(\phi)\lesssim\delta_c(\phi_p)$, where $\delta_c(\phi_p)=(\xi/2)/\ln(\sigma_0/\sigma_p)$
is the connectivity at the insulator-conductor transition.

\section{conclusions}
\label{concl}
In this article we have studied by theory and simulations the effects of short-range particle attractions
on the functional dependencies of the conductivity $\sigma$ of equilibrium dispersions of conducting
spherical fillers electrically connected through tunneling processes.
Instead of employing the traditional continuum percolation approach in which a fixed connectivity length
is introduced to mimic the extent of the tunneling electron transfer, we have solved the network equations
by treating each conducting particle as being connected to all others by tunneling conductances.
We have shown that the lack of a fixed connectivity distance gives rise to new and important features which
the continuum percolation approach fails to predict, and which could be crucial in the design of
nanocomposite materials with unique electrical properties.

In particular, we have found that the conductivity at intermediate and large contents of the conducting fillers
is extremely sensitive to the profile of the attractive potential, while the low-density regime is generally governed only
by the stickiness of the attraction. As a consequence, materials with similar values of the conductor-insulator
concentration $\phi_p$ may display at $\phi>\phi_p$ conductivities which differ by several orders of magnitude
if the attraction is tuned while keeping the stickiness constant.
More generally, we have demonstrated that the knowledge of $\phi_p$ and of its dependence on the attraction
are not sufficient conditions to presume the overall behavior of $\sigma$ as a function of $\phi$, in striking contrast
to the percolation transition approach represented by Eq.~\eqref{sigmaperc}.

As discussed in Sec.~\ref{sec:EMA} our simulation results can be reproduced by using a recently developed effective
medium theory which, for sufficiently small values of $\xi/D$, expresses the conductivity
in terms of the radial distribution function at contact, explicitly relating the different regimes of conduction
to the depth and range of the potential well. Furthermore we have shown how the critical path
approximation overcomes the limitations of the continuum percolation approach by allowing the
connectivity length to adjust for given $\phi$ and potential profile, so to represent the dominant
tunneling distances governing $\sigma$.

Although we are not aware of experiments on conducting colloidal composites where the electronic conductivity is 
tuned by controlling the filler attraction \cite{noteConcl}, our work may nevertheless stimulate experiments in
this direction.
In principle, our predictions may be systematically tested in experiments by changing the size and
concentration of depletants in colloidal conductors. In particular we have predicted that an optimal
enhancement of the conductivity can be realized by high concentrations of depletants with small
sizes compared to those of the conducting particles, so to drive the system towards the AHS limit.
Furthermore, stronger enhancements are expected in composites with conducting filler
sizes sufficiently large compared to the tunneling decay length $\xi$, so to avoid a smearing out
of the attraction potential well.

We conclude by pointing out that the system considered here, \emph{i.e.}, attracting fillers with
spherical shapes, has served primarily to illustrate the general principles at the basis of the interplay
between attraction and tunneling. Stronger effects of the attraction are expected when
the conducting particles have shape anisotropy, as in fibrous or plate-like fillers, where the mutual
orientations of the particles couple with the attraction interaction. This represents certainly an issue
of great interest for future investigations.

\begin{acknowledgements}
B. N. acknowledges support by the Swiss National Science Foundation (Grant No. 200020-135491).
M. A. M. is supported by EPSRC (U.K.).
\end{acknowledgements}


\begin{thebibliography}{50}

\bibitem{Bauhofer2009}
W. Bauhofer and J. Z. Kovacs, Compos. Sci. Technol. {\bf 69}, 1486 (2009).

\bibitem{Kuilla2010}
T. Kuilla, S. Bhadra, D. Yao, N. H. Kim, S. Bose, J. H. Lee,
Prog. Polym. Sci. {\bf 35}, 1350 (2010).

\bibitem{Grunlan2001}
J. C. Grunlan, W. W.  Gerberich, and L. F. Francis,
Polym. Eng. Sci. {\bf 41}, 1947 (2001).

\bibitem{Kyrylyuk2011}
A. V. Kyrylyuk, M. C. Hermant, T. Schilling, B. Klumperman, C. E. Koning, and
P. van der Schoot, Nat. Nanotechnol. {\bf 6}, 364 (2011).

\bibitem{Vigolo2005}
B. Vigolo, C. Coulon, M. Maugey, C. Zakri, and P. Poulin, Science {\bf 309}, 920 (2005).

\bibitem{Kharchenko2004}
S. B. Kharchenko, J. F. Douglas, J. Obrzut, E. A. Grulke, and K. B. Migler,
Nature Mater. {\bf 3}, 564 (2004).

\bibitem{Hermant2009}
M. C. Hermant, B. Klumperman, A. V. Kyrylyuk, P. van der Schoot, and C. E. Koning,
Soft Matter {\bf 5}, 878 (2009).

\bibitem{Stauffer1994}
D. Stauffer and A. Aharony, {\it Introduction to Percolation Theory}
(Taylor \& Francis, London, 1994).

\bibitem{Sahimi2003}
M. Sahimi, {\it Heterogeneous Materials I. Linear Transport and Optical Properties}
(Springer, New York, 2003).

\bibitem{Vionnet2005}
S. Vionnet-Menot, C. Grimaldi, T. Maeder, S. Str\"assler, and P. Ryser,
Phys. Rev. B {\bf 71}, 064201 (2005).

\bibitem{Balberg1984}
I. Balberg, C. H. Anderson, S. Alexander, and N. Wagner,
Phys. Rev. B {\bf 30}, 3933 (1984).

\bibitem{Otten}
R. H. J. Otten and P. van der Schoot, Phys. Rev. Lett. {\bf 103}, 225704 (2009);
J. Chem. Phys. {\bf 134}, 094902 (2011).

\bibitem{Chatterjee2010}
A. P. Chatterjee, J. Chem. Phys. {\bf 132}, 224905 (2010).

\bibitem{Kusy1977}
R. P. Kusy, J. Appl. Phys. {\bf 48}, 5301 (1977).

\bibitem{Johner2009}
N. Johner, C. Grimaldi, T. Maeder, and P. Ryser, Phys. Rev. E {\bf 79},
020104(R) (2009).

\bibitem{Chiew1983}
Y. C. Chiew and E. D. Glandt, J. Phys. A: Math. Gen. {\bf 16}, 2599 (1983).

\bibitem{Bug1985}
A. L. R. Bug, S. A. Safran, G. S. Grest, and I. Webman,
Phys. Rev. Lett. {\bf 55}, 1896 (1985); S. A. Safran, I. Webman, and G. S. Grest,
Phys. Rev. A {\bf 32}, 506 (1985).

\bibitem{Schilling2007}
T. Schilling, S. Jungblut, and M. A. Miller, Phys. Rev. Lett. {\bf 98}, 108303 (2007).

\bibitem{Kyrylyuk2008}
A. V. Kyrylyuk and P. van der Schoot, Proc. Natl. Acad. Sci. U. S. A. {\bf 105}, 8221 (2008).

\bibitem{Ambrosetti2010a}
G. Ambrosetti, C. Grimaldi, I. Balberg, T. Maeder, A. Danani, and P. Ryser,
Phys. Rev. B {\bf 81} , 155434 (2010).

\bibitem{Ambrosetti2010b}
G. Ambrosetti, I. Balberg, and C. Grimaldi, Phys. Rev. B {\bf 82}, 134201 (2010).

\bibitem{Nigro2011}
B. Nigro, G. Ambrosetti, C. Grimaldi, T. Maeder, and P. Ryser, Phys. Rev. B {\bf 83}, 064203 (2011).

\bibitem{Lekkerkerker2011}
H. N. W. Lekkerkerker and R. Tuinier, {\it Colloids and the Depletion Interaction}
(Springer, Dordrecht, 2011).

\bibitem{Baxter1968}
R. J. Baxter, J. Chem. Phys. {\bf 49}, 2770 (1968).

\bibitem{Noro2000}
M. G. Noro and D. Frenkel, J. Chem. Phys. {\bf 113}, 2941 (2000)

\bibitem{Foffi2006}
G. Foffi and F. Sciortino, Phys. Rev. E {\bf 74}, 050401(R) (2006).

\bibitem{Largo2008}
J. Largo, M. A. Miller and F. Sciortino, J. Chem. Phys. {\bf 128}, 134513 (2008).

\bibitem{Miller2004}
M. A. Miller and D. Frenkel, J. Chem. Phys. {\bf 121}, 535 (2004).

\bibitem{NoteSec3}
Note that in practice $\sigma_p$ may also be given by the lowest measurable
conductivity which depends on the experimental setup.

\bibitem{Grimaldi2011}
C. Grimaldi, EPL Europhys. Lett. {\bf 96}, 36004 (2011).

\bibitem{Yuste1994}
S. B. Yuste and A. Santos, J. Chem. Phys. {\bf 101}, 2355 (1994).

\bibitem{Largo2005}
J. Largo, J. R. Solana, S. B. Yuste, and A. Santos,
J. Chem. Phys. {\bf 122}, 084510 (2005).

\bibitem{CPApapers}
V. Ambegaokar. B. I. Halperin, and J. S. Langer, Phys. Rev. B {\bf 4}, 2612 (1971);
B. I. Shklovskii and A. L. Efros, Sov. Phys. JETP {\bf 33}, 468 (1971).

\bibitem{notesigma0}
The values of the prefactor $\sigma_0$ which best fit the MC data lie between
$0.08$ and $0.1$, \textit{i.e.}, they are almost independent of the potential profile.

\bibitem{noteConcl}
In Ref.~\cite{Vigolo2005} attraction between nanotubes was tuned by means of depletion interaction induced by 
micelles of ionic surfactants. Since the resulting solvent was a conductive electrolyte, the 
dc conductivity provided no useful information and only the effects of attraction on the dielectric 
constant were measured.

\end{thebibliography}
\end{document}